\documentclass[journal=jpcbfk,manuscript=article]{achemso}

\usepackage[version=3]{mhchem} 
\usepackage{xcolor}
\usepackage[normalem]{ulem}

\setkeys{acs}{maxauthors = 10}

\newcommand{\ryr}{\textcolor{black}}
\newcommand{\ryrb}{\textcolor{black}}

\author{Ruiyu Wang}
\affiliation{Institute for Physical Science and Technology, University of Maryland, College Park, MD 20742, USA}

\author{Shams Mehdi}
\affiliation{Biophysics Program, University of Maryland, College Park, Maryland 20742, USA}
\alsoaffiliation{Institute for Physical Science and Technology, University of Maryland, College Park, MD 20742, USA}

\author{Ziyue Zou}
\affiliation{Department of Chemistry and Biochemistry, University of Maryland, College Park, MD 20742, USA}

\author{Pratyush Tiwary}
 \email{ptiwary@umd.edu}
 \affiliation{Department of Chemistry and Biochemistry, University of Maryland, College Park, MD 20742, USA}
 \alsoaffiliation{Institute for Physical Science and Technology, University of Maryland, College Park, MD 20742, USA}

\title[title]
  {Is the Local Ion Density Sufficient to Drive NaCl Nucleation \ryr{from the Melt and Aqueous Solution}?}

\begin{document}

\begin{tocentry}
    
    \includegraphics{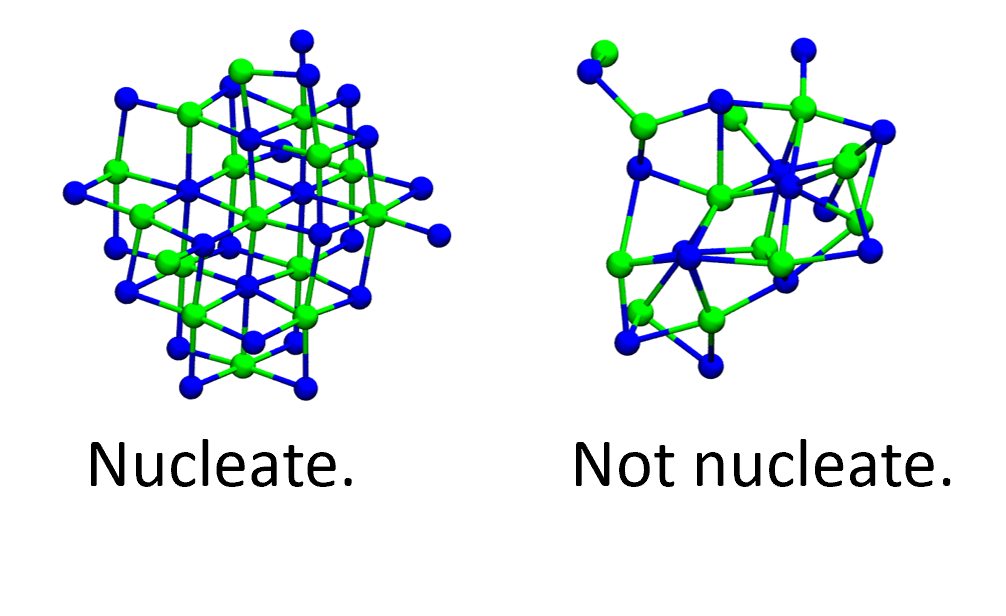}
    \centering
\end{tocentry}


\begin{abstract}
\begin{sloppypar}

Even though nucleation is ubiquitous in different science and engineering problems, investigating nucleation is extremely difficult due to the complicated ranges of time and length scales involved. In this work, we simulate NaCl nucleation in both molten and aqueous environments using enhanced sampling all-atom molecular dynamics with deep learning-based estimation of reaction coordinates. By incorporating various structural order parameters and learning the reaction coordinate as a function thereof, we achieve significantly improved sampling relative to traditional ad hoc descriptions of what drives nucleation, particularly in the aqueous medium. Our results reveal a one-step nucleation mechanism in both environments, with reaction coordinate analysis highlighting the importance of local ion density in distinguishing solid and liquid states. However, while fluctuations in the local ion density are necessary to drive nucleation, they are not sufficient. Our analysis shows that near the transition states, descriptors such as enthalpy and local structure become crucial. Our protocol proposed here enables robust nucleation analysis and phase sampling, and could offer insights into nucleation mechanisms for generic small molecules in different environments.
\end{sloppypar}
\end{abstract}

\section{Introduction}\label{Introduction}

Understanding the forces driving crystallization is relevant from a fundamental scientific perspective and for a wide range of practical applications.\cite{sosso_cr_2016,kapil_nat_2022,slater_nrc_2019,sosso_jcp_2016,montero_jcp_2023,cheng_pnas_2019} In natural environments, crystalline solids are major components of rocks in the form of feldspar containing Si, Al, O and other ions\cite{chardon_esr_2006,yuan_esr_2019,wang_wcms_2021,wang_jcis_2022,radha_pnas_2010} with implications for climate change. 
In industrial scenarios, understanding crystallization processes is essential for the improvement of material design and synthesis.  for instance, Lithium deposition has a significant impact on the lifetime and behavior of batteries;\cite{yang_nc_2023,chen_am_2021}  the nucleation and growth of metal-organic frameworks happens in solutions but the mechanism is not clear\cite{kollias_jacs_2022}; and active pharmaceutical ingredients also require specific crystal polymorphs.\cite{mortazavi_cc_2019,metz_cgd_2022,zou_jpcb_2021,price_csr_2014}

The nucleation process is the first step to form small nuclei from melt or solution, which then undergoes a growth stage to form larger crystals.\cite{erdemir_acr_2009} However, investigating such nucleation processes is extremely limited using experimental techniques because of the difficulty observing nanoscale details involving just dozens of atoms. Further complicating the issue is the variety of temporal scales - nucleation might take hours or days to occur, but when it does happen it can occur as quickly as milliseconds or significantly faster.\cite{nakamuro_jacs_2021}  One longstanding popular approach to model nucleation is the Classical Nucleation Theory (CNT). CNT assumes a single free energy barrier given as a function of the radius $r$ of spherically shaped nuclei clusters. It describes nucleation as a competition between the free energy cost for forming a surface (scaling as $r^2$) and the free energy benefit from a new crystalline phase (scaling as $r^3$).\cite{finney_chemrxiv_2023} CNT is an oversimplified model and completely overlooks the ``two-step'' mechanism in addition to other assumptions, often leading to significant errors in the nucleation rate even for simple systems.\cite{tsai_jcp_2019,diemand_jcp_2013} As a result, it is not surprising that CNT is not able to describe the nucleation of other molecules with more complicated interactions or shapes, such as urea and glycine.\cite{zou_pnas_2023,zhao_arxiv_2023}

\begin{figure}[!ht]
    \begin{center}
    \includegraphics[width=0.95\textwidth]{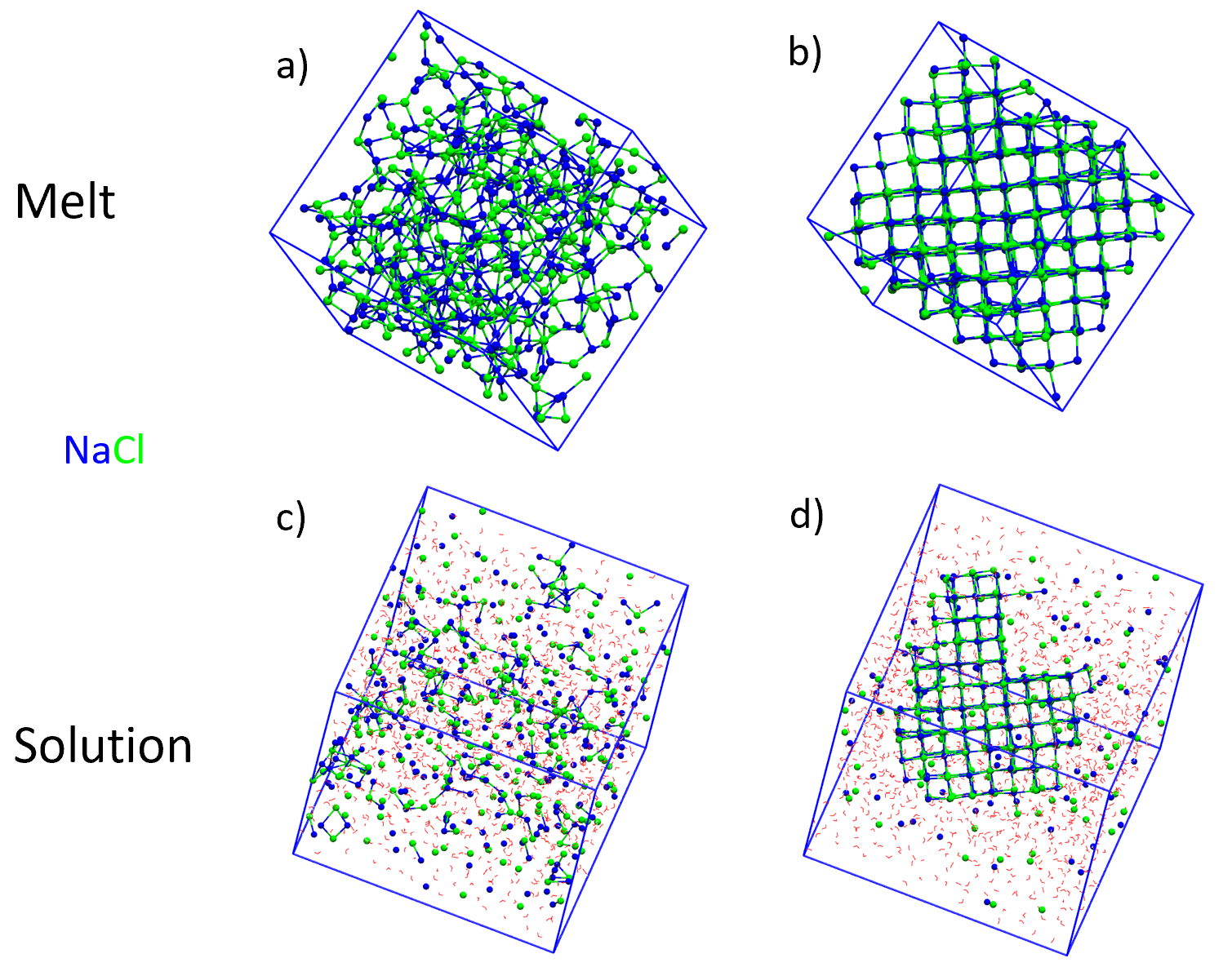}
    \caption{Snapshots of NaCl structures in simulations. a) Melt, b) solid in the melt, c) solution, d) solid in the solution. \ce{Na+} and \ce{Cl-} atoms are blue and green, respectively. Sticks in snapshots represent bonds between \ce{Na+} and \ce{Cl-}, if $d_{Na-Cl}<0.33$ nm. Simulation box is shown in blue lines. The snapshots are generated using VMD.\cite{vmd}}    
    \label{fig:nacl}
    \end{center}
\end{figure}

In this work, we study the 
\ryr{NaCl nucleation from both the melt and aqueous solution.\cite{nakamuro_jacs_2021,bian_acsami_2022,cedeno_jcp_2023,lamas_pccp_2021} We are interested in understanding whether the classical picture, that fluctuations in local ion density are the key driver of crystal nucleation, indeed holds true.\cite{r1c2,r1c3} Such insight is useful for a fundamental insight into nucleation and is hard to obtain through existing experimental or simulation techniques. By understanding what is it that really drives nucleation, in NaCl and more generally in other systems, we hope to develop pathways to truly controlling the process of nucleation and propensity for the same for complex engineering applications.} 
The liquid and solid structure of the melt (Fig.~\ref{fig:nacl}a and b) and solution of (Fig.~\ref{fig:nacl}c and d) NaCl are illustrated in respective figures. 
NaCl is a simple crystal with the face-centered cubic structure comprising the two species \ce{Na+} and \ce{Cl-}. There are 6 directly contacted counterions around each ion,
forming an octahedral geometry. 
The nucleation of NaCl from aqueous solution has been extensively studied through  experimental and simulation approaches.\cite{nakamuro_jacs_2021,bian_acsami_2022,cedeno_jcp_2023,anwar_angew_2011,jiang_jcp_2018_c,jiang_jcp_2019,chakraborty_jpcl_2013,finney_fd_2022,karmakar_jctc_2019,zimmermann_jcp_2018, giberti_iucrj_2015,giberti_jctc_2013,zahn_prl_2004,hwang_cs_2021} In aqueous solution, nucleation mechanism of NaCl is concentration dependent: at low concentration, it agrees with the CNT;\cite{jiang_jcp_2018} at high concentration, it is believed to follow the 2-step mechanism.\cite{finney_fd_2022,jiang_jcp_2019} NaCl nucleation from the melt remains equally if not more debated.\cite{Badin_prl_2021} 
Single-step and multi-step nucleation mechanisms are distinguished by the number of energy barriers during crystallization.  Both mechanisms require ion aggregation as the prerequisite of nucleation, forming a region with high local ion density.\cite{erdemir_acr_2009} Whether the resulting ion cluster is ordered or amorphous determines if it is a single- or multi-step mechanism respectively. The amorphous precursors with high local liquid density have been observed in experiments for ice and zeolites.\cite{vekilov_cgd_2004,kumar_cm_2016} Nevertheless, how exactly local density affects the nucleation process is still not completely clear. 

To study the atomic scale determinants of nucleation processes, an alternative to experiments and CNT is molecular dynamics (MD) simulations. MD allows studying the microscopic picture during nucleation processes because its resolution is at the atomic level and on a time scale of fs to ns.  However, the timescale of nucleation processes could be seconds or much slower, which is inaccessible for MD.\cite{diemand_jcp_2013,salvalaglio_jcp_2016} To overcome such limitations, biased enhanced sampling has been introduced that accelerates rare events. Of particular interest to us is well-tempered metadynamics (WTMetaD), which adds a time-dependent bias potential along a low-dimensional degree of freedom to escape free energy minima.\cite{bussi_nrp_2020,barducci_prl_2008,barducci_wcms_2011} While in principle such an approach is guaranteed to converge in some limit of time for any biasing variable,\cite{dama_prl_2014}
in practice it is best if the biasing is performed along the reaction coordinate (RC) or even an approximation of it.\cite{bussi_nrp_2020}
An approximation can be learned through a range of methods including artificial neural network based methods,\cite{tiwary_pnas_2016} such as state predictive information bottleneck (SPIB)\cite{wang_jcp_2021} which we use here. These methods\cite{mehdi_arxiv_2022,eric_cossms_2023} generally speaking learn the RC as a linear or non-linear combination of a bigger list of structural Order Parameters (OPs) that can distinguish different competing phases.\cite{zou_pnas_2023,zou_jpcb_2021} \ryr{One advantage is that SPIB is a semi-supervised method, which requires relatively less human intervention to propose input OPs.}

In this work we investigate the nucleation of NaCl from melt and in the aqueous medium through enhanced sampling simulations with deep learning based estimation of RCs. We start with collating several structural OPs and learn RCs as linear combinations of OPs. By performing biased sampling simulations along the approximate RC, we obtain significantly enhanced sampling of NaCl nucleation. The improvement in sampling is especially striking for NaCl in water, outperforming sampling achieved with non-deep learning based biasing variables. 

We find that at the temperature and concentration studied in this work, nucleation from both the melt and aqueous solution follow the one-step mechanism.
We analyze the RC further in terms of what really drives the process of nucleation over transition states both in melt and solution.\cite{mehdi_arxiv_2022}
Overall the learned RCs have contributions from the coordination number of ions, which clearly distinguish the solid and liquid states. However, when we focus our attention on interpreting the deep neural network RCs in the vicinity of the transition states, we find that OPs other than the density descriptor coordination number show high importance. We find that the energy-based descriptor enthalpy and local structural descriptors are key to driving the system over the nucleation barrier in the melt and solution, respectively.
Although forcing ion aggregation can help form the NaCl solid for both melt and aqueous solution, high local ion density is insufficient to describe the transition of the nucleation process,
unlike CNT, whose only order parameter is the size of the precursor nuclei, which is also a descriptor of the volume of high density region. 
\ryr{Since CNT takes the size of crystalline structure as the order parameter, it is not surprising that in one-step mechanism, ion aggregation and ordered packing happen at the same time during the nucleation process.}
The protocol used in this work provides a robust, automated means for studying nucleation of small molecules and demonstrates the ability to sample different phases, estimate free energy surfaces and reveal the nucleation mechanism. Such a protocol should also be applicable to the study of different crystal-forming systems.

\section{Methods}\label{methods}
\subsection{Order parameters (OPs)}
Many OPs have been developed and used to study nucleation. In this work, we use 3 different classes of OPs: 
ion local density descriptors related to coordination number, 
ion local order descriptor related to Steinhardt bond OPs and the purely thermodynamic descriptor enthalpy.

\subsubsection{Coordination numbers related OPs}
The coordination number (CN) is defined as the number of ions within the cutoff distance. In this work, we calculate the CN only considering Na-Cl pairs. In other words, the CN of a \ce{Na+} is how many \ce{Cl-} are around it and vice versa. A continuously differentiable form of the CN is calculated through
\begin{equation}
  s(i)=\sum_j \frac{1-(r_{ij}/r_0)^6}{1-(r_{ij}/r_0)^{12}}
\end{equation}
where $s(i)$ is the coordination number of atom i, j is the atom index of each counterion, $r_0=0.33$ nm is the cutoff and $r_{ij}$ is the distance between atom i and j. To reduce the collection of CN values for all atoms to a few scalar quantities, we calculate (i) the average of all CNs, (ii) the second moment
of all CNs, (iii-iv) number of ions with coordination number more than 4 or 5. These are are selected as OPs and are labelled $\overline{N}$, $\mu_N^2$, $N_{4+}$ and $N_{5+}$, respectively. The ``more\_than'' OPs (iii)-(iv), as implemented in PLUMED\cite{plumed} are calculated by:

\begin{equation}
  N_{s_0+}=\sum_i [1-\sigma(s_i)]
\end{equation}
\begin{equation}
  \sigma(s_i)=\frac{1-(s_i/s_0)^6}{1-(s_i/s_0)^{12}}
\end{equation}
where $s_0$ is 4 or 5 for (iii) or (iv) respectively. Such a continuously differentiable CN may not be equal to the exact CN which is strictly an integer, but it captures the trend of ion pairing.

Coordination number is a descriptor of the local density. It has been used to study nucleation in different systems, such as Argon, since high coordinated Argon atoms belong to liquid phase and lower ones are gas.\cite{tsai_jcp_2019,moroni_prl_2005}  The NaCl crystal is the octahedral rock-salt structure, indicating that both \ce{Na+} or \ce{Cl-} have a coordination number of 6, whereas liquid (both melt and solution) has a lower CN. However, $\overline{N}$ is not enough to distinguish some structures, such as there are more than one cluster (including the solid-liquid co-existence) and defects on crystals. That is why $\mu_N^2$ is introduced.\cite{salvalaglio_jcp_2016,ten_jcp_1998} $N_{4+}$ and $N_{5+}$ describe number of ions with high coordination number and they can be used to distinguish the blob structure and solid.

\subsubsection{Steinhardt bond OPs}
These OPs are descriptors of ion local order of packing and describe the symmetry of atoms around a central atom. In this work, we only consider Na-Cl directly contacted ion pairs, similar to the definition of coordination number. The OPs are calculated as:
\begin{equation}
  q_{lm}(i) = \frac{\sum_j \sigma(r_{ij})\mathbf{Y_{lm}(r_{ij})}}{\sum_j \sigma(r_{ij})}
\end{equation}
where $\mathbf{Y_{lm}}$ is the $l^{th}$ order spherical harmonic and m ranges from -l to l.\cite{steinhardt_prb_1983} In this work, we select the $4^{th}$ and $6^{th}$ order Steinhardt OPs and their average values are biased (marked as $\overline{q4}$ or $\overline{q6}$).  These OPs have been widely used to study the phase transition of Lennard-Jones particles,\cite{moroni_prl_2005} ice\cite{gasser_science_2001} and NaCl\cite{jiang_jcp_2018,sun_crystals_2020}. Other moments are not considered in this work because the computational cost of calculating these OPs on-the-fly during molecular dynamics can be prohibitively high.

\subsubsection{Enthalpy}
Enthalpy is a thermodynamical descriptor of the system which has been shown to be useful in the study of nucleation without prior knowledge of the symmetry of the phases that may form.\cite{piaggi_prl_2017} It is calculated as:
\begin{equation}
  H=E+pV
\end{equation}
where $H$, $E$, $p$ and $V$ denote enthalpy, potential energy, pressure and volume of the full system respectively.  The enthalpy OP was first introduced to study the phase transition of metal crystals.\cite{piaggi_prl_2017}

\subsection{Simulation settings}
All MD simulations were carried out using GROMACS 2022.3\cite{gromacs_1995,gromacs_2015} All simulations were performed using the constant number, pressure and temperature (NPT) ensemble using a time step 2 fs 
The pressure was maintained at 1 bar using isotropic
Parrinello-Rahman barostat with a relaxation time of 10 ps.\cite{npt2} The temperatures of the melt and aqueous solution systems were maintained at 1125 and 300 K respectively using velocity scaling with a relaxation time of 0.1 ps.\cite{nvt} There were 256 Na-Cl pairs in all simulations and 1350 water molecules, corresponding to a concentration of 8.86 mol/L, which is around 1.5 times of the saturated concentration (denoted as $1.5c_s$). We used the force field described in Ref  \cite{zimmermann_jacs_2015}, which predicts the rock salt polymorph and avoids the artifact of the wurtzite polymorph.  Water OH bonds were fixed using the LINCS algorithm.\cite{lincs}  The cutoff of short-range interactions was 1 nm and long-range electrostatic interactions are calculated using particle-mesh Ewald summations.\cite{ewald}  All simulations were carried out for 500 ns, except as explained later 
that the initial and production trajectories in solution were of 900 ns to obtain better convergence. 

WTMetaD simulations were carried out with PLUMED 2.8.1.\cite{plumed,bonomi_nm_2019} For simulations using both 1d and 2d RCs, bias potential with initial height 5 kJ/mol were added to simulations every 2 ps. The bias factor was set as 100.

\section{Results and discussions}\label{results}

\subsection{State Predictive Information Bottleneck}

SPIB is applied in this work to extract machine learning based RCs from MD simulations. SPIB is a variant of the Reweighted Autoencoded Variational Bayes method\cite{ribeiro_jcp_2018} and its details have been discussed before.\cite{wang_jcp_2021} In summary, SPIB learns RCs through the encoder of a variational autoencoder following the information bottleneck, compressing the trajectory to a low-dimensional representation in the latent space but containing maximal targeted information about future state of the system after a certain time-delay. 
SPIB is able to distinguish different metastable states, such as liquid and solid states in this work.
The type of the encoder can be linear or nonlinear.
When a linear encoder is selected, SPIB will learn RCs as a linear combination of selected OPs of a simulation, which can be biased and enhance the sampling in WTMetaD simulations. Nonlinear RCs have higher expressive capacity than the linear ones, as a result, they are applied when analyzing the final trajectories in order to distinguish metastable states. Hyperparameters used in SPIB are listed in the Table~S1 and S2. In this work, linear RCs are used to bias in WTMetaD simulations and non-linear RCs are used to identify metastable states and transition states.

\subsection{RC for melt nucleation is not the same as for nucleation in water}
\begin{figure}[!ht]
    \begin{center}
    \includegraphics[width=0.75\textwidth]{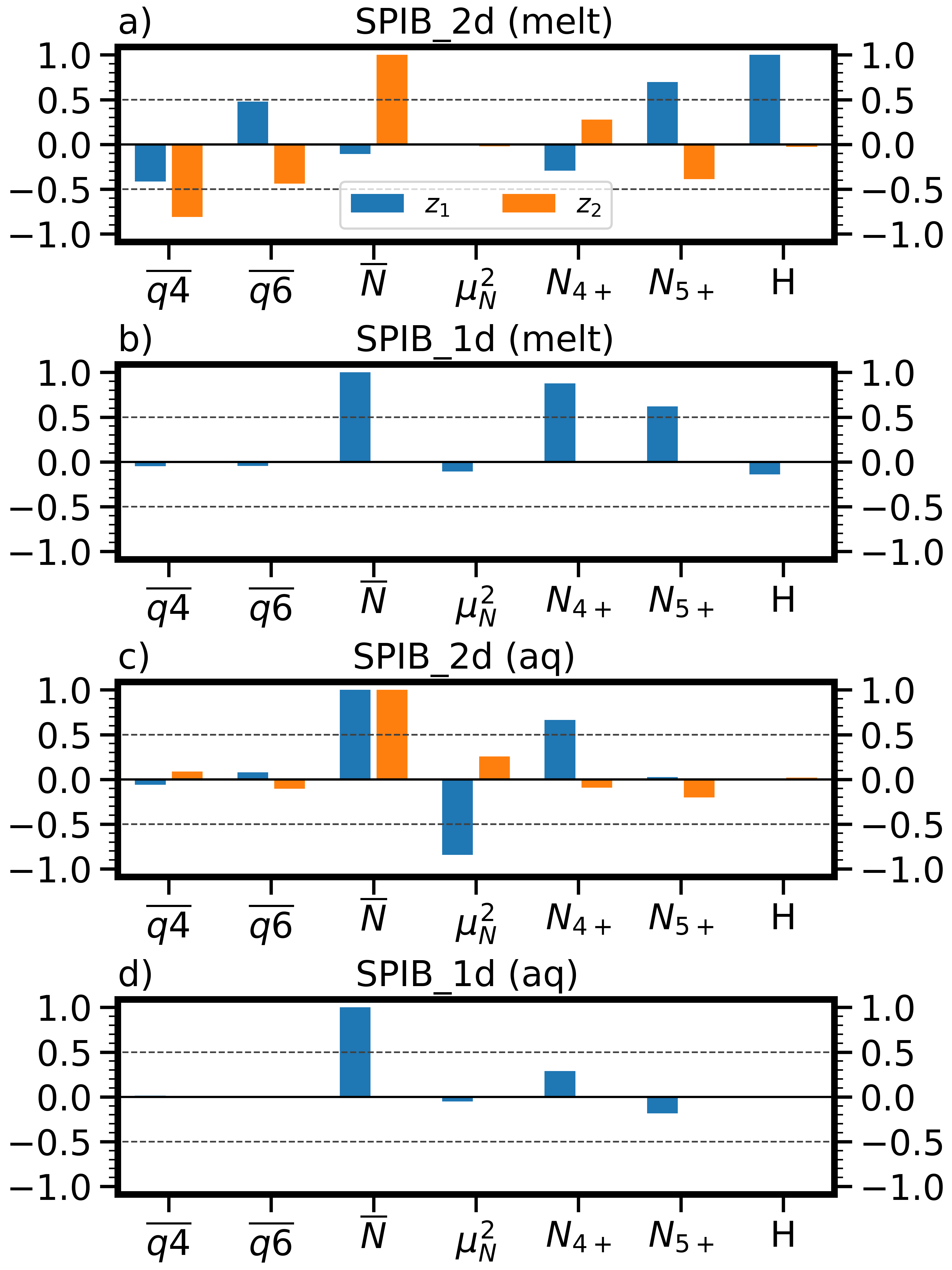}
    \caption{Scaled coefficients of RCs learned by SPIB and used in the production run. ``melt'' and ``aq'' represent simulations of the melt (a,b) or aqueous solution (c,d). ``1d'' and ``2d'' represent the dimensionality of RCs. The two components in 2d RCs are marked as $z_1$ and $z_2$. }
    \label{fig:coef}
    \end{center}
\end{figure}

Following the procedure discussed in the section \textit{protocol to run AI augmented sampling} in SI, using a linear encoder for SPIB,\cite{mehdi_jctc_2022} we obtain the scaled coefficients of RCs (Fig.~\ref{fig:coef}) such that maximum absolute value for any component of the RCs is 1. In all obtained RCs, $\overline{N}$ has the highest coefficient, indicating that it is the most efficient OP to distinguish the solid and liquid states for nucleation from both the melt (Fig.~\ref{fig:coef}a and b) and the solution (Fig.~\ref{fig:coef}c and d). We generate both 2d (Fig.~\ref{fig:coef}a and c) and 1d RCs (Fig.~\ref{fig:coef}b and d) because sampling along high dimensional RCs reveals better free energy surface, but its high computational cost may not be acceptable. 
One could argue that since for both systems the training data through the trial trajectory (marked as SPIB0) is obtained by biasing $\overline{N}$, it is possible that SPIB memorizes $\overline{N}$ overlooking other OPs, and thereby overestimating its relevance. We show in Fig.~S4 that this is not the case. There we report an SPIB result using a trajectory biasing $H$ and $\overline{N}$ still has high weight. 

While the average coordination number $\overline{N}$ has the highest weight across all schema for both nucleation from melt and from solution, the RCs also have non-trivial contributions from other OPs.  As discussed in the section \textit{order parameters} in SI, $\overline{N}$ describes ion aggregation and high $\overline{N}$ means more contact ion pairs and high ion local density. Some coefficients are significantly different between nucleation from melt and from solution, indicating different underlying mechanisms. We first highlight $N_{4+}$ and $N_{5+}$, the number of ions with coordination number greater than 4 and 5 respectively. An ion with this many counterions is highly likely to belong to a crystalline solid instead of amorphous ion clusters. $N_{4+}$ and $N_{5+}$ play essential roles in the RCs for nucleation from melt and solution, respectively. 
In the melt, the difference of $\overline{N}$ between liquid and solid is smaller than that in aqueous solution (Fig.~\ref{fig:nacl}, S6 and S7). 
The liquid of the melt has an $\overline{N}$ of 3 but about 1 in the solution. As a result, $N_{4+}$ is enough to distinguish the liquid and solid states in solution but $N_{5+}$ is necessary for the melt.
The relative high weight of $N_{4+}$ and $N_{5+}$ in RCs indicates that $\overline{N}$, the local density of ions, is not sufficient to fully describe the nucleation process. Here the number of ions belonging to solid phase is a necessary OP.
The second difference is in the weight of the enthapy $H$. Low $H$, corresponding to low energy represents solid structures.\cite{piaggi_prl_2017}
In simulations of melt, $H$ carries a high weight in the $z_1$ component of the RC $z_1$ 
in SPIB\_2d (Fig.~\ref{fig:coef}a). However, the contribution of $H$ is nearly zero for nucleation from solution, as here $H$ includes contributions from both ions and water thereby having poor signal-to-noise ratio from the perspective of nucleation. 
In addition, Steinhardt OPs show non-zero weights in the 2d RCs of only for nucleation from melt (Fig.~\ref{fig:coef}a and c). This can also be seen from the MD trajectories, where Steinhardt OPs shows a clear trend in distinguishing liquid and solid states in the melt, whereas in the solution the effect is less pronounced (Fig.~S6 and S7).

The last difference comes from the dimensionality of the RCs. 
Ideally, coefficients of 2d RCs are supposed to be as orthogonal as possible to take full advantage of OPs. We find that in simulations of both melt and solution, the highest sampling efficiency (discussed below) is achieved using 2d RCs whose two components have low similarity (Fig.~S2).

\subsection{Increased sampling efficiency}
We expect that biasing approximate RCs learnt through SPIB should show improved sampling efficiency. This can be roughly defined as the number of back-and-forth phase transitions observed per simulation time. Unlike the straightforward simulation time in unbiased simulations,
in biased simulations such as metadynamics, the effect of the bias has to be taken into account, which can be done approximately through the accelerated time in the qualitative spirit of infrequent metadynamics.\cite{tiwary_prl_2013} We define this accelerated time in the SI (Eq.~S4), and denote it as $t_{acc}$. Roughly speaking this accounts for the number of transitions seen per unit of biasing work done on the system \cite{tiwary_jpcb_2015}, with the idea being that biasing along a improved variable should show more transitions between metastable states of interest.

\begin{figure*}[!ht]
\begin{center}
    \includegraphics[width=0.95\textwidth]{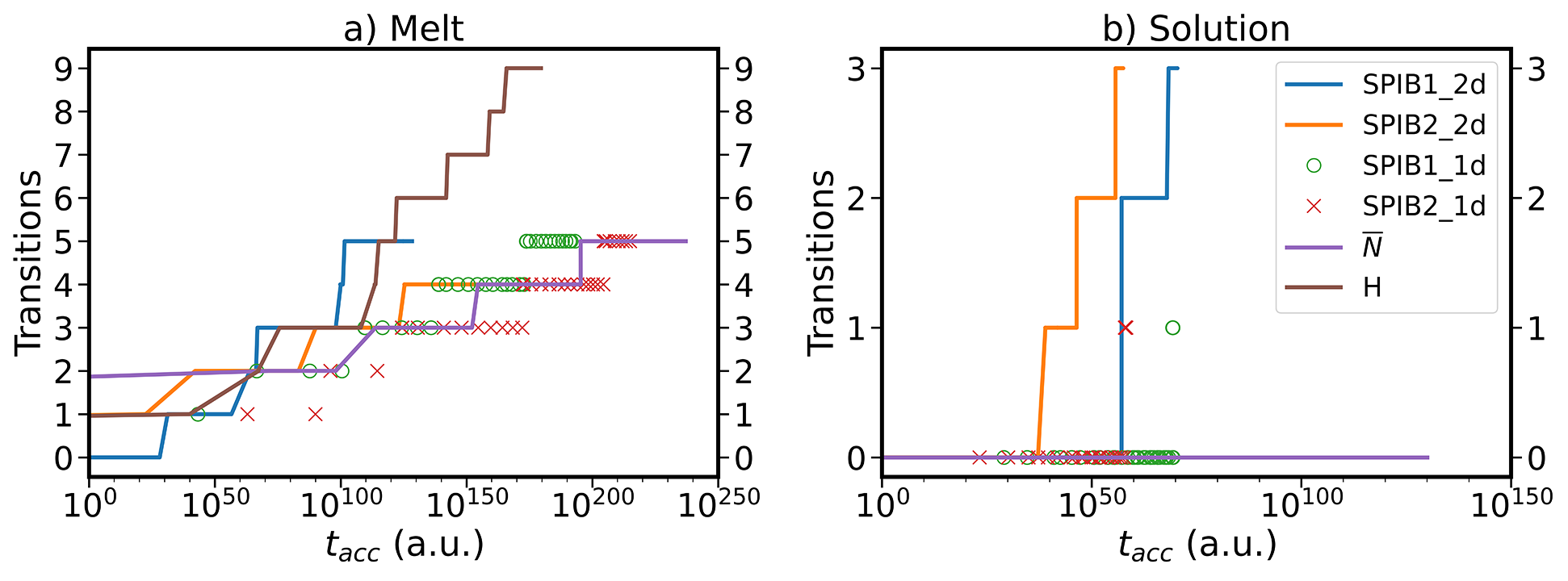}
    \caption{Number of phase transitions vs. accelerated time $t_{acc}$ (defined in Eq.~S4 in SI) for simulations of the melt (a) and solution (b). Only the first 500 ns of the simulation time is plotted. The legend in (b) shows what is biased in the plots for both melt and solution. Biasing only $H$ in simulations of solution is not tested and there is no brown curve in plot (b).}
    \label{fig:transitions}
\end{center}
\end{figure*}

A phase transition is defined by the change between liquid and solid states, that is, either L-S or S-L is counted as 1 transition and L-S-L is counted as 2 transitions. For simulations from the melt, in the first 500 ns simulations time, at least 5 phase transitions are observed (Fig.~\ref{fig:transitions}a).
Simulations biasing the enthalpy outperforms all other approximations to the RC. This is in agreement with previous observations that driving the fluctuation of enthalpy can improve the sampling between liquid and crystal phases.\cite{piaggi_prl_2017} However, as discussed above, $H$ cannot be applied to simulations of solutions. 
The fact that more frequent phase transitions are observed when using 2d RCs for nucleation from both the melt and solution shows the importance of the dimensionality of RCs.

A similar trend is also observed in simulations of nucleation from solution (Fig.~\ref{fig:transitions}b). First, increased sampling efficiency is observed when using high dimensional RC and phase transition frequency shows 2d RC$>$1d RC$> \overline{N}$. If biasing $\overline{N}$ only, no phase transition happens in the first 500 ns simulation time. 
Although the general trend is that more SPIB cycles increase the sampling efficiency, the SPIB2\_2d RC of the melt is an exception because the two components in the RC are too similar and make the sampling efficiency comparable to 1d RCs (Fig.~S3). As a result, we use SPIB1\_2d in the production run.
By balancing the cost of training and efficiency improvement, we use the RCs of SPIB2\_2d for production runs. The increased number of phase transitions is essential evidence that SPIB learned RCs enhance the sampling of simulations, driving the nucleation and dissolution of NaCl in both melt and solution.

\subsection{Evaluation of order parameter importance in reaction coordinates}

An additional question regarding SPIB RCs is to evaluate the importance of the OPs. 
Although SPIB can analyze the whole trajectory and extract a linear combination of all OPs to distinguish the liquid and solid states, such model is not necessary to drive the phase transition, because it may not match the committor for the nucleation process.\cite{bolhuis_arpc_2002} As a result, the weights of SPIB RCs are not equivalant to the importance of OPs that drive the phase transition.
To appropriately investigate system behavior near the transition state, we trained a SPIB model with non-linear encoder on the final trajectories. Afterwards, we employ a recently developed artificial intelligence (AI) model interpretation technique, Thermodynamically Explainable Representations of AI and other black-box Paradigms\cite{mehdi_arxiv_2022} to identify feature relevance in the vicinity of the transition state.
The procedure to carry out TERP analysis is discussed in the SI.

\begin{figure*}[!ht]
    \begin{center}
    \includegraphics[width=0.95\textwidth]{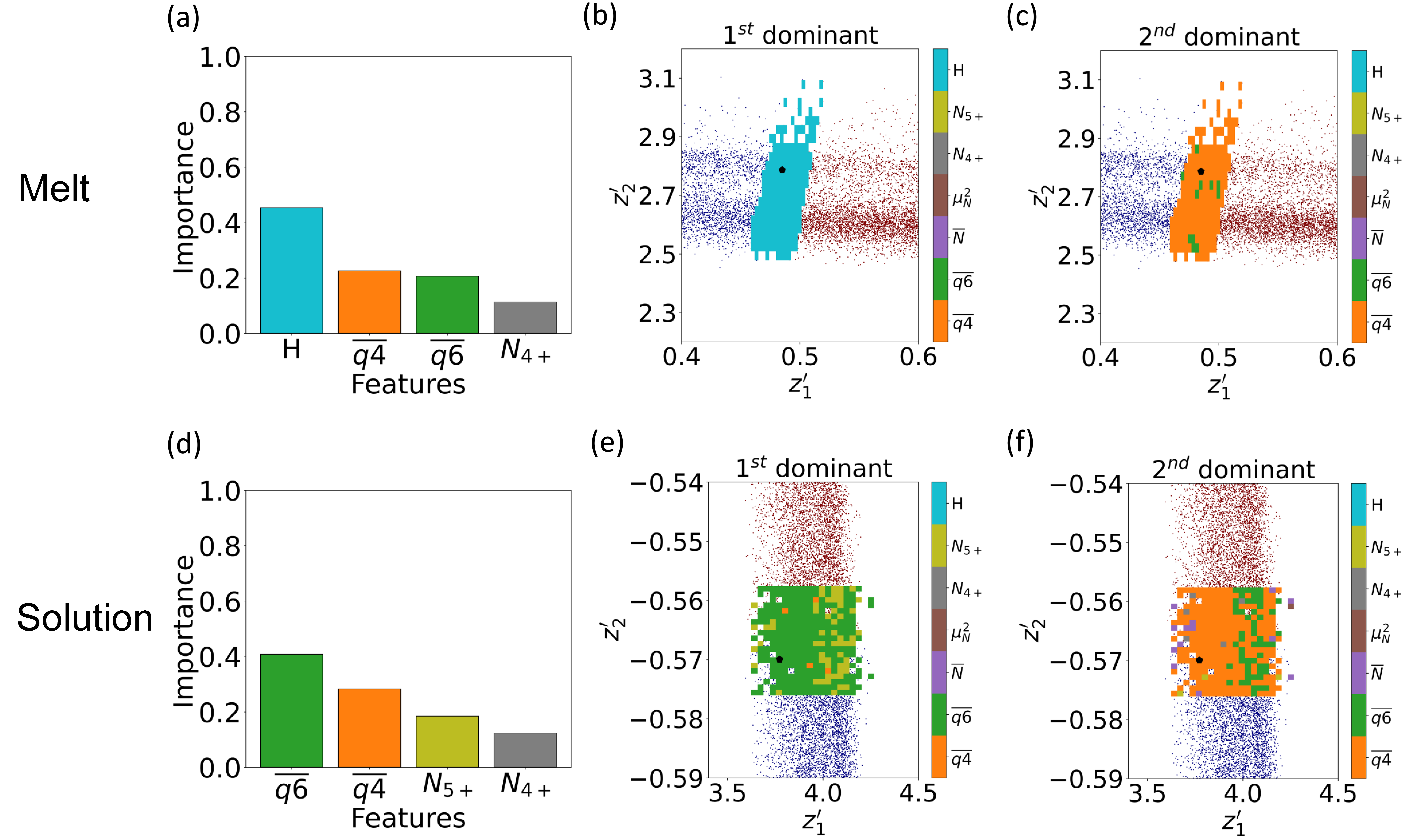}
    \caption{OP importance analysis using TERP. (a)-(c) and (d)-(f) are results for simulations of the melt and solution, respectively. (a) and (d) The TERP result of a selected point; for these specific points TERP optimal model predicts four non-zero coefficients only.
    (b)-(f) The dominant features at the transition state predicted by TERP. The $1^{st}$ and $2^{nd}$ dominant features show that most points in that grid predict that feature as the highest and $2^{nd}$ highest importance. The black points are the example points and their TERP feature importance are plotted in (a) and (d). The background red and blue points are SPIB learned non-linear RCs.}
    \label{fig:terps}
    \end{center}
\end{figure*}

In a nutshell, TERP ranks the local importance of each feature for each selected point. Features with high importance drive the nucleation of NaCl. 
For instance, Fig.~\ref{fig:terps}a shows the TERP ranking of a select point in the transition state region for nucleation from melt. The transition state region itself is ascertained from the boundary between metastable states classified by SPIB.\cite{wang_jcp_2021,beyerle_jpcb_2022} The OPs $H$, $\overline{q4}$ and $\overline{q6}$ show their importance in driving nucleation at that point.
In addition to explaining one configuration, here we study a more broader transition state ensemble. 
The transition states are divided into many grids based on non-linear RCs $z'_1$ and $z'_2$. We run TERP for all points in each grid and the $n^{th}$ dominant feature is determined if TERP predicts rank that feature as the $n^{th}$ importance for most points in the grid.

For nucleation from the melt, Fig.~\ref{fig:terps}b and c show that TERP predicts $H$ and $\overline{q4}$ as the first and second most important feature of most points in each grid. Nearly all grids are light blue in Fig.~\ref{fig:terps}b, indicating that there is only one nucleation mechanism for the melt that described by the OP $H$. The observation is consistent with the fact that biasing $H$ leads to the best phase transition sampling efficiency. 
$\overline{q4}$ is relatively important after $H$, as the descriptor of local ion structures (Fig.~\ref{fig:terps}c) and has been used in previous simulations of melt NaCl.\cite{valeriani_jcp_2005} Interestingly, $\overline{N}$ disappears though it has the highest weight in SPIB learned RCs. Our explanation is that $\overline{N}$ is a descriptor of ion local density and enough to distinguish liquid and solid states. However, simply compressing ions together results in a unstructured blobs instead of ordered packing solid.

For nucleation from solution, TERP predicts that $\overline{q6}$, $\overline{q4}$ and $N_{5+}$ are important for the example point (Fig.~\ref{fig:terps}d) in the transition state ensemble. We also find that $H$ is not important at all as discussed in the previous section. 
$\overline{q6}$ is the most important (Fig.~\ref{fig:terps}e) and $\overline{q4}$ is the $2^{nd}$ most important (Fig.~\ref{fig:terps}f) during nucleation from the solution, demonstrating the importance of local ion ordered structure.\cite{sun_crystals_2020} 
A small fraction of members of the transition state ensemble requires the OP $N_{5+}$, ions with high CN, which is an approximation of ions belonging to the solid structure as discussed before.\cite{finney_fd_2022}
Similar to nucleation from the melt, the weight of $\overline{N}$ is trivial near the transition state. Such TERP results show that ion local density does not drive the nucleation of NaCl near the transition state, but the local ordered structure of ions and the number of ordered ions that belong to the solid structure should not be neglected.

Fig.~\ref{fig:terps} shows that in simulations of both melt and solution, though $\overline{N}$ has the highest coefficients in SPIB learned RCs, near the transition state, other OPs in the RCs also have high importance to drive the nucleation of NaCl. The fact that multiple OPs have a relative high feature score is consistent with the fact that 2d RC has better sampling efficiency than 1d RC.

\subsection{Sampling, free energy and nucleation mechanism}
\begin{figure}[!ht]
    \includegraphics[width=0.95\textwidth]{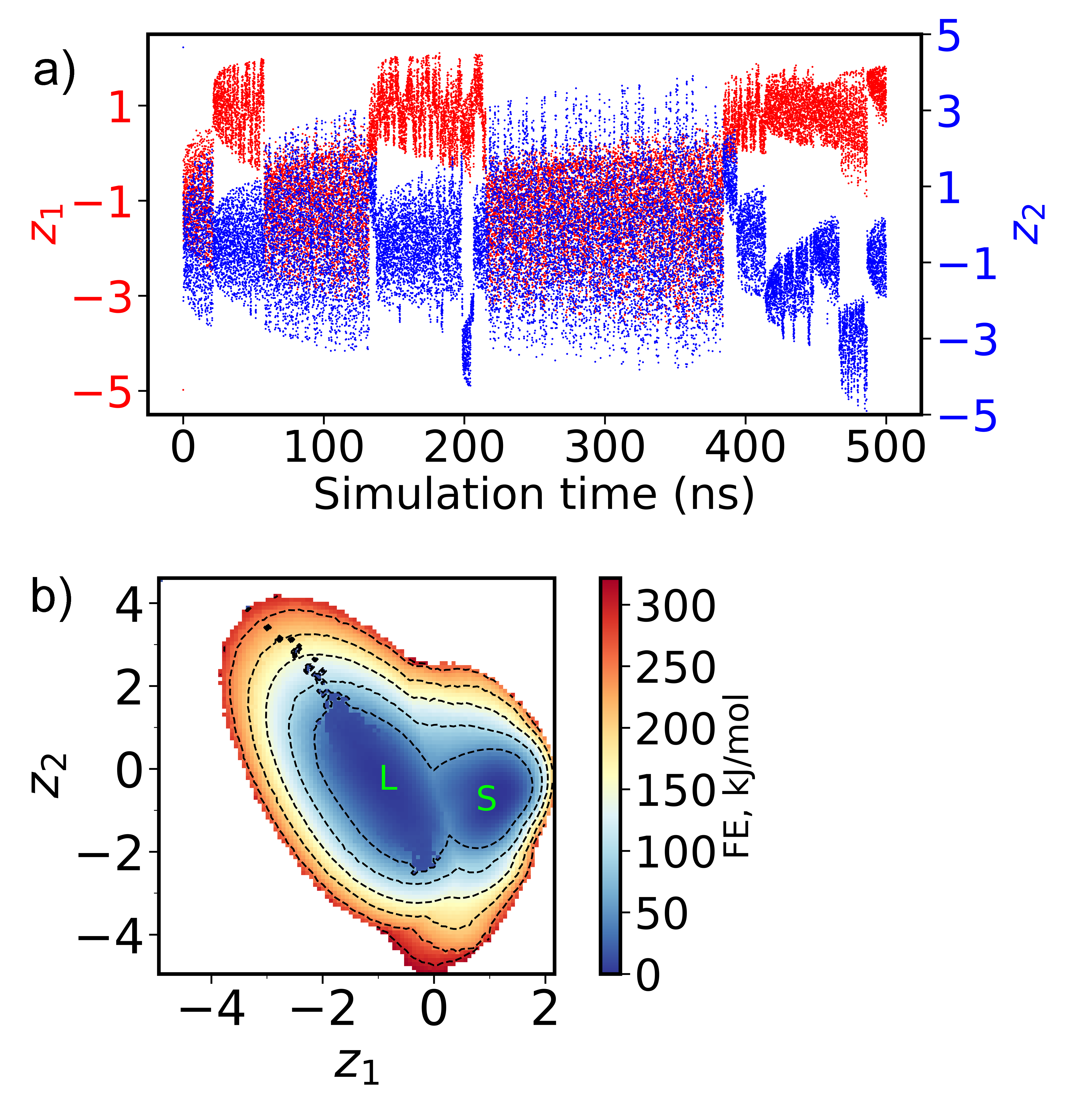}
    \caption{Simulation results of WTMetaD of the melt NaCl. (a) Trajectory represented using RC $z_1$ and $z_2$. (b) FES of the melt calculated using reweighted histogram method. Label ``L'' and ``S'' represent the liquid and solid structures in the RC space. In all plots, RC $z_1$ and $z_2$ is unitless; the unit of FES and error as indicated on the color bar is kJ/mol. The contour lines is every 50 kJ/mol and ``FE'' is short for ``free energy''.}
    \label{fig:femelt}
\end{figure}

In the simulations from the melt, the time series trajectory using SPIB1\_2d RC shows 2 states clearly in Fig.~\ref{fig:femelt}a, especially for $z_1$, whose high value represents solid and vice versa. The fluctuation of $z_2$ also contains information of phase transition: small fluctuation happens when ions are solid. Interestingly, $z_2$ reveals more information of solid structures. In the last of 100 ns of the simulation, defects are observed in the solid and several small metastable states at $z_2$ = 0, -1, -2 and -4 (Fig.~\ref{fig:femelt}a) represent solid structures with variant defects.


\begin{sloppypar}  

The free energy surface (FES) calculated using the reweighted histogram method\cite{tiwary_jpcb_2015} (Eq.~S5), is consistent with the observed trajectory.  There are two states on the FES. As discussed above, the left and right free energy minima represent liquid and solid, respectively (Fig.~\ref{fig:femelt}b). There are several noise points on the FES of the liquid state but they can be neglected. The value of the FES does not show too much difference between liquid and solid states, but the barrier is about 50 kJ/mol, requiring enhanced sampling methods to get across it.  Although in the RC space, the area of the left state is larger, the calculated free energy change from liquid to solid is $\Delta G_{L-S}=-11.24 \pm 0.95$ kJ/mol. Unfortunately, at 1125 K NaCl is liquid experimentally and a positive free energy change is expected. Due to less time spent in the transition state ensemble, sampling there may not be sufficient and WTMetaD simulations may not be converged in that area. Another factor could be the force field used in this work, which is designed for solutions and may not predict the correct melting point of NaCl. Last, the solid and liquid states are distinguished using SPIB with tuned hyperparameters (discussed in SI), but those states are not perfected states observed in experiments, such as variant defects of solid structures in the simulations. Those factors are considered the reason for the discrepancy between experiments and simulations. 

\end{sloppypar}

\begin{figure}[!ht]
    \includegraphics[width=0.95\textwidth]{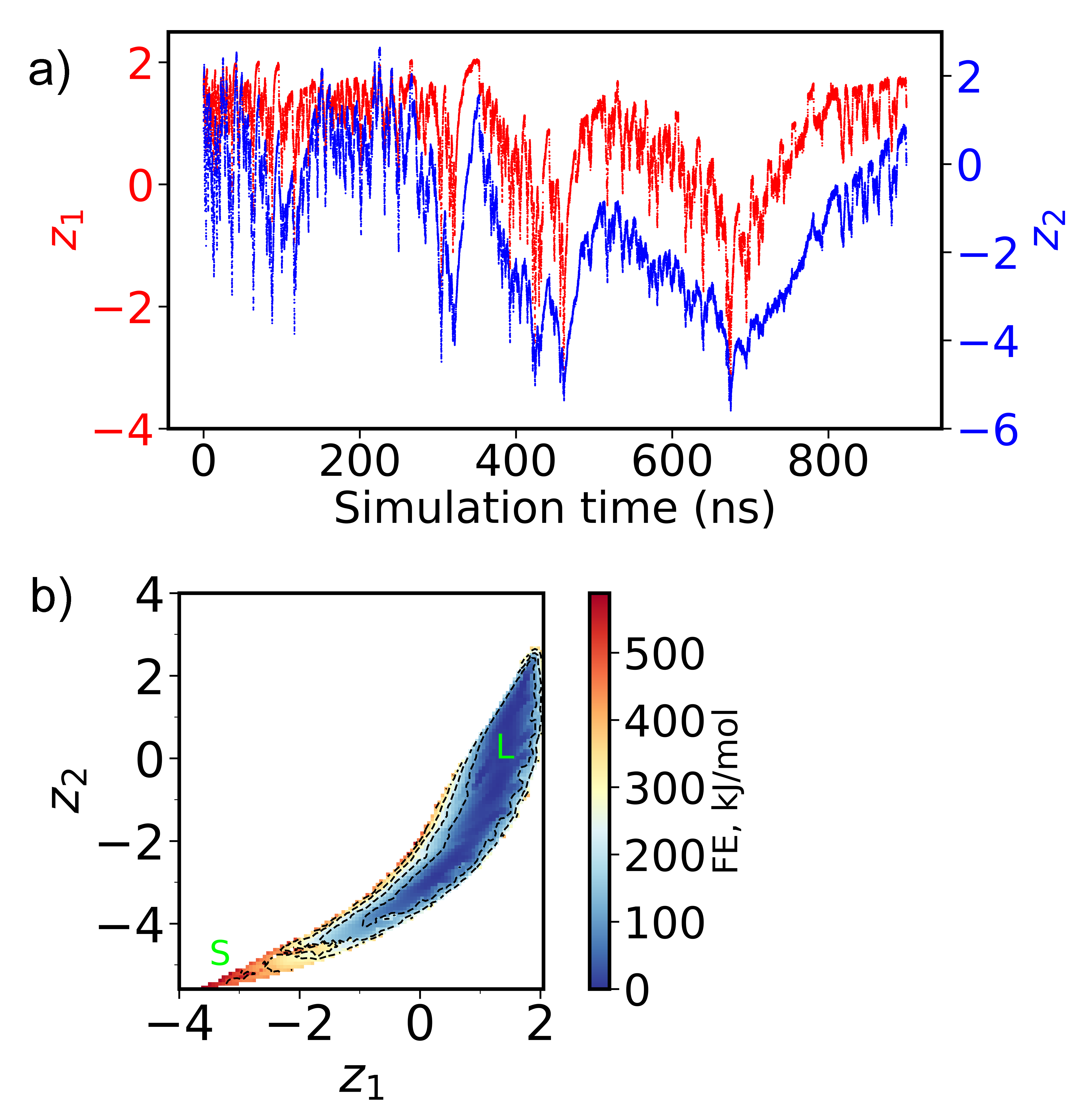}
    \caption{Simulation results of WTMetaD of the NaCl solution. (a) Trajectory represented using RC $z_1$ and $z_2$. (b) FES calculated using the same method as the melt. Label ``L'' and ``S'' represent the liquid and solid structure in the RC space.  In all plots, RC $z_1$ and $z_2$ is unitless; the unit of FES and error as indicated on the color bar is kJ/mol. The contour lines is every 100 kJ/mol and ``FE'' is short for ``free energy''.}
    \label{fig:feaq}
\end{figure}

Unlike simulations of the melt, the first challenge in analyzing simulations of the solution is to clearly distinguish the states of nucleation from solution, for instance, the unstructured blob and solid states. The trajectory along the 2d RC shows that transitions of these states are continuous without a clear boundary (Fig.~\ref{fig:feaq}a). The low values of $z_1$ and $z_2$ represent the solid state, and in the RC space, the left bottom and right top corners represent solid and liquid states, respectively. Since the similarity of $z_1$ and $z_2$ is about 0.45, $z_1$ and $z_2$ are more linearly correlated in simulations of solution than that of melt (Fig.~\ref{fig:feaq}b).

The FES for nucleation from the solution (Fig.~\ref{fig:feaq}b) is consistent with the trajectory. There is only one metastable state on the FES plot, corresponding to the liquid 
\ryr{state, and there is no observable barrier on the FES plot. However, it does not mean that the barrier does not exist. In addition, the free energy difference of liquid and solid configurations is about 500 kJ/mol,}
demonstrating the necessity of WTMetaD to enhance the sampling because solids may never appear in brute force simulations. The free energy change is $\Delta G_{L-S}=8.34 \pm 6.3$ kJ/mol, indicating that the liquid is more stable. Besides the convergence of simulations discussed above, an additional difficulty to estimate the free energy in the simulations is the identification of states. In the simulations,
\ryr{due to the depletion of ions when solid forms, the critical size of the solid cannot be accurately defined.}

The FES plots, when projected along SPIB-learnt reaction coordinates, are consistent with the proposed nucleation mechanism discussed above. There is one barrier in the FES of the melt, consistent with the one-step mechanism observed from the trajectories. Furthermore, the liquid and solid states are on the left and right side on the FES plot, parallel to $z_1$ (Fig.~\ref{fig:femelt}b). Since the weight of $H$ in $z_1$ is dominant, both FES and TERP predict enthalpy as the most important OP that drive the phase transition of the melt. As for nucleation in aqueous solution, the fact that no barrier on the FES makes multi-step nucleation mechanism less possible. The nucleation process shows two segments on the FES plot. From the top right to ($z_1=0$,$z_2=-3$), $z_2$ dominates the process, corresponding to the aggregation of ions with increasing $\overline{N}$. From the point to the bottom left, $z_1$ with OPs besides $\overline{N}$ is needed for the growth of the NaCl nuclei. Although TERP highlights Steinhardt bond OPs that do not contribute to SPIB learned RCs, the relative high weights of $N_{4+}$ in $z_1$ still proves that OPs that describe local environment order besides local density of ions are needed to drive the nucleation of NaCl in aqueous solution.

\subsection{Comparison with previous studies}

The nucleation process is affected by many factors, such as temperature in melt, concentration in solution, as well as force field, enhanced sampling methods and the size of simulation box in both simulations. Temperature is crucial in nucleation from the melt. While the experimentally reported melting point of NaCl is 1073 K, some simulations report 1064 K.\cite{anwar_jcp_2002} Simulations with lower temperatures will definitely favor the solid state.\cite{valeriani_jcp_2005} In this work, at 1125 K, liquid is expected as the equilibrium state but the free energy from our simulation shows slightly negative $\Delta G$ for liquid-solid transition. The discrepancies arise from the factors mentioned above. 

We observe 1-step nucleation mechanism in this work but discussions on the mechanism of the melt in previous work is limited.\cite{Badin_prl_2021,valeriani_jcp_2005}
The nucleation of NaCl in aqueous solution is concentration dependent.\cite{finney_chemrxiv_2023} Simulations with high concentration leads to 2-step nucleation mechanism, forming an amorphous aggregation blob then ordered solid.\cite{finney_chemrxiv_2023,finney_fd_2022} In this work, NaCl concentration is low ($1.5c_s$) and 1-step mechanism is consistent with previous reports.

\ryrb{In the framework of CNT, the reaction coordinate is the size of crystals, equivalent to the number of crystalline ions.  However, such RC is not bias-able and not applicable in enhanced sampling MD simulations, but SPIB learned RCs can overcome such limitation.}

Last but not the least, finite-size effects significantly affect the free energy of nucleation by depletion of the solution. The depletion reduces the concentration of the solution and overestimates the free energy of the solid state.\cite{li_jpcl_2023} The issue can be fixed by applying constant chemical potential MD with a control region to maintain the concentration of the solution, though the simulation settings are more complicated.\cite{karmakar_jctc_2019}

\section{Conclusion}\label{conclusion}

In this work, we have investigated the nucleation of NaCl from both the melt and aqueous solution using machine learning-based enhanced sampling molecular dynamics simulations.
We propose order parameters that describe ion local order, ion local density and the energy of simulations. AI is applied to extract RCs as a linear combination of OPs and significantly enhances the sampling of phase transition between the liquid and solid NaCl. By evaluating the importance of OPs at the transition state, 
we find that the local density of ions could distinguish the solid and liquid states but is not enough to drive the nucleation. Energy and local order descriptors are necessary for the nucleation from both the melt and aqueous solution, respectively. 
We observed the one-step nucleation mechanism for both the melt and solution, which is consistent with previous works. The free energy surfaces of the phase transition are also comparable with previous literature and support the nucleation mechanism observed in this work.
This work provides an automated paradigm to study nucleation of salts. Combination of machine learning and enhanced sampling can help drive the nucleation and evaluate how fundamental physical properties affect the phase transition and reveal the mechanism of nucleation process. We expect that the workflow developed in this work will be applied to the nucleation of more complicated molecules with practical applications. \ryrb{Investigating nucleation processes of small organic molecules can provide physical insights on synthesis of materials or drug molecules, which is beneficial to experimental chemists as well.\cite{kollias_jacs_2022,sosso_cr_2016} }


\begin{acknowledgement}
This research was entirely supported by the US Department of Energy, Office of Science, Basic Energy Sciences, CPIMS Program, under Award DE-SC0021009. We are grateful to NSF ACCESS Bridges2 (project CHE180053) and University of Maryland Zaratan High-Performance Computing cluster for enabling the work performed here.

\end{acknowledgement}

\begin{suppinfo}

Simulation details, additional analysis can be found in the supporting information.

\end{suppinfo}

\section{Note}
The authors declare no competing financial interest.

\section{Data availability}
\begin{sloppypar}
The input files necessary to reproduce simulations in this work are available on GitHub (simulation settings) and PLUMED NEST (WTMetaD) at https://github.com/ruiyuwangwork/NaCl\_nucleation\_melt\_aq and https://www.plumed-nest.org/eggs/23/036/. 

The data that support the findings of this study are available from the corresponding author upon reasonable request.
\end{sloppypar}

\bibliography{achemso-demo}

\end{document}